\documentclass[pdflatex,sn-mathphys-num]{sn-jnl}% Math and Physical Sciences Numbered Reference Style

\usepackage{graphicx}%
\usepackage{multirow}%
\usepackage{amsmath,amssymb,amsfonts}%
\usepackage{amsthm}%
\usepackage[mathscr]{eucal}%
\usepackage[title]{appendix}%
\usepackage{xcolor}%
\usepackage{textcomp}%
\usepackage{manyfoot}%
\usepackage{booktabs}%
\usepackage{algorithm}%
\usepackage{algorithmicx}%
\usepackage{algpseudocode}%
\usepackage{listings}%
\theoremstyle{thmstyleone}%
\theoremstyle{thmstyletwo}%

\theoremstyle{thmstylethree}%

\begin{document}

\title[Semi-Automatic Correction of 3D Tubular Structure Skeletons]{Semi-Automatic Correction of 3D Tubular Structure Skeletons via Component-Wise MST and Filtered Delaunay Triangulation}

%%=============================================================%%
%% Author information
%%=============================================================%%

\author[1,2,3]{\fnm{Ruoxuan} \sur{Yang}}\email{ruoxuan.yang@telecom-paris.fr}

\author*[4,5,6]{\fnm{Chuan} \sur{Li}}\email{chuan.li@sorbonne-universite.fr}

\affil[1]{\orgname{Shanghai Jiao Tong University}, \orgaddress{\city{Shanghai}, \country{China}}}

\affil[2]{\orgname{Télécom Paris}, \orgaddress{\city{Palaiseau}, \country{France}}}

\affil[3]{\orgname{Institut Polytechnique de Paris}, \orgaddress{\city{Palaiseau}, \country{France}}}

\affil*[4]{\orgname{Sorbonne Université}, \orgaddress{\city{Paris}, \country{France}}}

\affil[5]{\orgname{LIPADE, Université Paris Cité}, \orgaddress{\city{Paris}, \country{France}}}

\affil[6]{\orgname{Télécom SudParis, Institut Polytechnique de Paris}, \orgaddress{\city{Palaiseau}, \country{France}}}


\abstract{Skeletonization of tubular structures from 3D imaging is essential for tasks such as morphometric analysis, transport or flow simulation, and procedural planning in domains including vascular networks, plant root systems, and neural connectomes. However, automatic skeleton extraction often introduces topological artifacts, such as erroneous connections between nearby branches and fragmented centerlines caused by noise or missing data. Correcting these artifacts manually can be time-consuming and error-prone, especially when precise interaction is required.

We present a semi-automatic correction method that reconstructs a plausible centerline connection from minimal user input. Given a user-selected source and target point, our method traces a path by combining (i) component-wise minimum spanning trees for stable local propagation and (ii) a filtered 3D Delaunay edge graph for bridging gaps and handling ambiguous junctions. Candidate steps are ranked using a score that accounts for direction continuity, spatial proximity, component consistency, and target-directed progress. The output is an ordered polyline (or edge sequence) that can be used as a suggested correction and integrated into downstream skeleton post-processing workflows.

We implement the system in C++ with an interactive viewer based on Libigl and demonstrate representative qualitative results on brain vessel datasets, including correction of typical ``crossing'' and ``dotted'' artifacts. While our current validation is qualitative, the method is lightweight and serves as a practical building block toward more comprehensive interactive correction pipelines in biomedical imaging and related domains.}




\maketitle

\section{Introduction}\label{sec1}

\begin{figure}
  \centering
  \includegraphics[width=\textwidth, trim=0 50 0 30, clip]{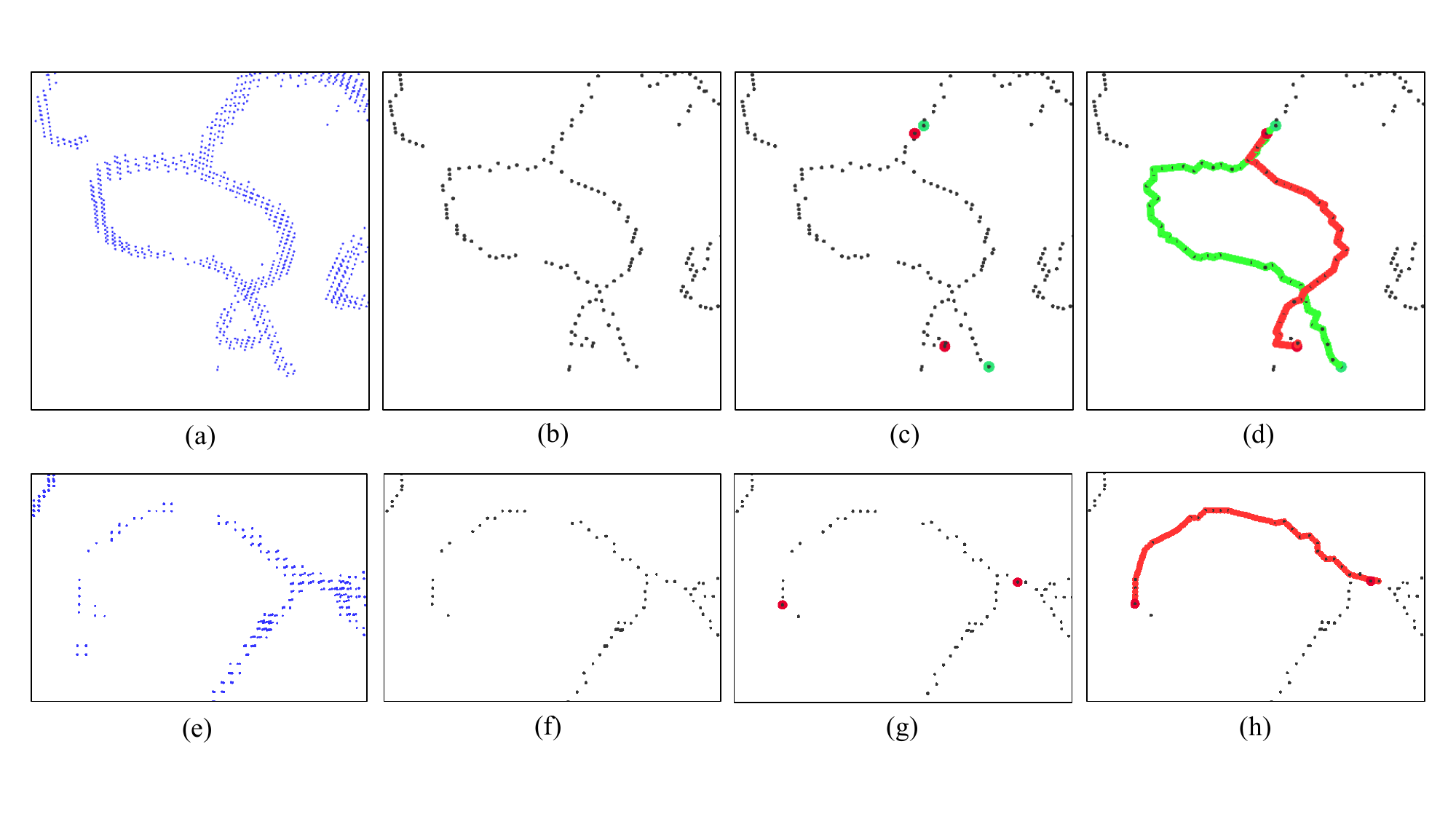}
  \caption{Typical correction scenarios on a brain-vessel point cloud. Top row: crossing segments. Bottom row: dotted segments. (a,e) Original point clouds; (b,f) extracted skeletons; (c,g) user-selected source--target pairs; (d,h) reconstructed connections. Similar artifacts also arise in other tubular structures, such as plant roots and neuronal arbors.}
  \label{fig:teaser}
\end{figure}

\begin{figure}[htbp!]
  \centering
  \includegraphics[width=\linewidth, trim=250 30 250 15, clip]{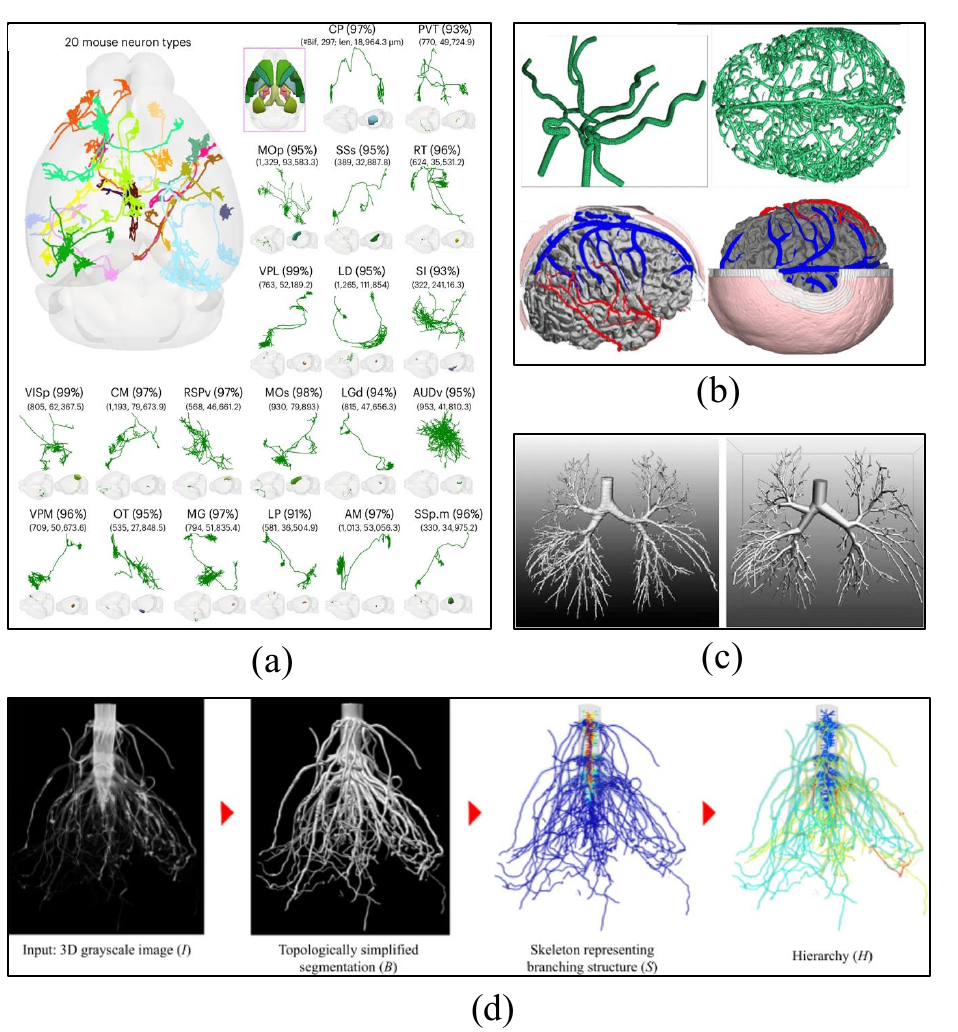}
  \caption{Some examples of tubular structures (a) Reconstruction of complete mouse neurons \cite{Zhang2024NatMethods_CAR}. (b) Segmentation of human brain vessel \cite{ghaffari2015automatic}. (c) Bronchial tree models \cite{gemci2008computational}. (d) Segmentation of maize root \cite{zeng2021toporoot}. }
  \label{fig:tubulars}
\end{figure}

Computer technologies are integral to modern science and engineering, enabling analysis in fields ranging from medicine to agriculture and neuroscience. A key area of research across these domains is the extraction and analysis of 3D tubular structures. These include, for example, blood vessels \cite{kirbas2004review} and bronchial trees \cite{selvan2016extraction} in medical imaging, plant root systems \cite{zeng2021toporoot, jiang2019three} and tree architectures \cite{raumonen2013fast} in botany, and the intricate networks of neuronal arbors within neural connectomes \cite{dorkenwald2023cave, dorkenwald2024neuronal} (see Figure~\ref{fig:tubulars}). Owing to their narrow, elongated geometries, these structures are commonly represented as centerline skeletons. Such representations facilitate diverse downstream tasks, including fluid dynamics or nutrient transport simulation \cite{antiga2008image, olufsen2000numerical}, morphometric analysis in vascular branching \cite{bullitt2003measuring}, root system architecture \cite{siddiqui2021genetics}, or neuronal morphology \cite{dorkenwald2023cave}, and guidance for interventions \cite{krissian2000model}.

%   \centering
%   \includegraphics[width=\textwidth, trim=0 50 0 30, clip]{fig/headfig.pdf}
%   \caption{Typical correction scenarios on a brain-vessel point cloud. Top row: ``crossing segments.'' Bottom row: ``dotted segments.'' (a,e) Original point clouds; (b,f) extracted skeletons; (c,g) user-selected source--target pairs; (d,h) reconstructed connections. Similar artifacts also arise in other tubular structures, such as plant roots and neuronal arbors.}

A wide range of automatic methods have been proposed to extract such skeletons from volumetric imaging data—such as MRA for vessels, CT or optical scans for plant roots, and electron microscopy for neural connectomes. However, these algorithms often struggle with noisy, incomplete, or ambiguous inputs, which are common in medical scans, complex soil environments for root imaging, and dense neural tissues. This challenge is further amplified in large-scale datasets typical of modern biomedical research, agricultural phenotyping, and connectomics, where manual validation and correction remain major bottlenecks \cite{nigolian2019invaner, bucksch2014image, dorkenwald2023cave}. To address these limitations, interactive correction tools have been developed, enabling users to manually refine or adjust algorithmic outputs. Although such tools provide flexibility and intuitive control \cite{nigolian2019invaner}, they can be tedious to use due to the demand for precise input, and in some cases, may rely excessively on manual interventions—potentially resulting in anatomically inconsistent reconstructions.

In this work, we address two common failure modes in tubular structure skeletonization: erroneous connections between distinct segments (referred to as “crossing segments”, as shown in Figure~\ref{fig:teaser} (a, b)) and fragmented segment representations resulting from data loss or noise (referred to as “dotted segments”, as shown in Figure~\ref{fig:teaser} (e, f)). We introduce a semi-automatic correction framework that, given minimal user input (e.g., source and target points), leverages the original geometric structure along with simple topological heuristics to infer and reconstruct anatomically plausible paths. Our goal is to reduce the burden of manual correction while maintaining the structural integrity of the underlying tubular structure.
% \paragraph{Why two endpoints?}
\medskip
\noindent\textbf{Why two endpoints?}
Many existing correction interfaces require users to trace the desired centerline by dragging across fragments or by placing multiple waypoints. While flexible, such interactions can be tedious and sensitive to missed clicks, particularly in cluttered 3D scenes where fragments are densely interleaved. Our design targets a complementary point in the interaction--automation trade-off: users specify the intended connectivity with a minimal source--target pair, and the system proposes a data-consistent connection that follows the local geometry whenever possible. We explicitly focus on two frequent failure modes---spurious cross-connections and fragmented segments---where the intended correction is typically well specified by endpoints. We also discuss cases where additional guidance is needed (e.g., highly ambiguous junctions) and outline extensions such as optional hint points in Section~\ref{sec:fail}.

\section{Related Work}
% \subsection{Volumetric Data for Tubular Structures}
% Volumetric data for tubular structures like vascular networks, plant roots, and neuronal connectomes essentially creates 3D digital maps of these intricate systems. For vascular networks, imaging techniques like CT or MRI scans generate slice-by-slice data, often stored in DICOM for clinical use or NIfTI for research, detailing vessel pathways and dimensions. For plant roots, data from X-ray CT, MRI, or optical scans are frequently translated into formats like RSML (Root System Markup Language), which describes the root's branching structure, geometry (like lengths and diameters), and growth over time. For neuronal connectomes, detailed 3D neuron shapes from light or electron microscopy are often stored in SWC files, representing the neuron's branching as a tree of connected points with coordinates and radii. More complex models including physiological properties use NeuroML , while massive raw image datasets from electron microscopy are handled by formats like HDF5 or N5.
\subsection{Volumetric Data for Tubular Structures}
Tubular structures such as vascular networks, plant roots, and neuronal arbors are typically extracted from volumetric imaging modalities including CT, MRI, optical scans, and electron microscopy. After segmentation, these structures are commonly represented as centerline skeletons stored in formats such as SWC or derived graph representations. Our method operates on skeleton point clouds and is agnostic to the specific volumetric storage format.

\subsection{Centerline extraction and skeletonization}
Centerline extraction and skeletonization are critical for analyzing 3D tubular structures, reducing complex geometries to simpler representations that preserve topological and geometric properties. A foundational method was proposed by Lee et al. \cite{Lee1994}, introducing an iterative thinning algorithm that symmetrically erodes surface layers while preserving topology, using a decision-tree framework for efficiency. This method is implemented in libraries such as ITK and scikit-image, and serves as the preprocessing step in our pipeline.

Alternative methods include Voronoi-based techniques such as the Power Crust \cite{Amenta2001}, which operates on point clouds, and contraction-based methods like Mean Curvature Skeletons \cite{Tagliasacchi2012}, which rely on curvature flow for producing smooth and robust skeletons. The variety of available approaches reflects different application requirements, with no universally optimal solution. Maintaining topological consistency remains a fundamental challenge in skeletonization.

\subsection{Graph-based path searching}
Graph-based algorithms have been extensively studied due to their versatility across a broad range of applications. Classical algorithms such as Kruskal’s \cite{Kruskal1956} and Prim’s \cite{Prim1957} for computing Minimum Spanning Trees (MSTs), Delaunay triangulation \cite{Delaunay1934} for spatial partitioning, and shortest-path algorithms like Dijkstra’s \cite{Dijkstra1959} and A* \cite{Hart1968} form the foundation for more advanced, task-specific methods in pathfinding and connectivity analysis.

These algorithms have been successfully applied to skeleton-related tasks. For example, MSTs are commonly used for pruning and simplifying complex skeletal graphs while preserving essential connections, as demonstrated in 3D plant modeling \cite{Dobbs2024Grapevine}. Delaunay triangulation and Voronoi diagrams play key roles in medial axis generation \cite{Ogniewicz1992Voronoi}. Shortest-path algorithms like Dijkstra’s and A* are widely used for tracing vessel centerlines in both images and skeletal graphs, including in interactive tools such as \cite{Abeysinghe2009Interactive}.

\subsection{Interactive correction tools}
Although automatic skeletonization techniques have advanced significantly, complex or noisy biological data often produce errors that require human correction. Interactive tools harness human perceptual strengths to refine automatically generated skeletons. For example, Abeysinghe and Ju \cite{Abeysinghe2009Interactive} proposed a method in which users define 3D skeleton curves by clicking 2D endpoints on volume views, inspired by the “intelligent scissors” paradigm. Their system also supports scribbling directly on the image to guide or correct curve generation.

Comprehensive editing toolkits such as SkeletonLab \cite{BARBIERI201623} provide users with operations to prune extraneous branches, merge nodes, and resample skeletal structures for uniformity, enabling application-specific adjustments. 

Our proposed method is closely related to the work of Nigolian et al. \cite{nigolian2019invaner}, who developed two interactive correction strategies for addressing artifacts such as “kissing vessels” and “dotted vessels.” Their system allows users to manually connect segments by clicking and dragging. While intuitive, this approach can be cumbersome—missing a single point may require restarting the process. In contrast, our method infers the connecting path automatically based on only two user-specified endpoints, streamlining the correction workflow.

\section{Technical Details}
\subsection{Data}
\begin{figure*}[htbp!]
  \centering
  \includegraphics[width=\linewidth, trim=75 50 80 90, clip]{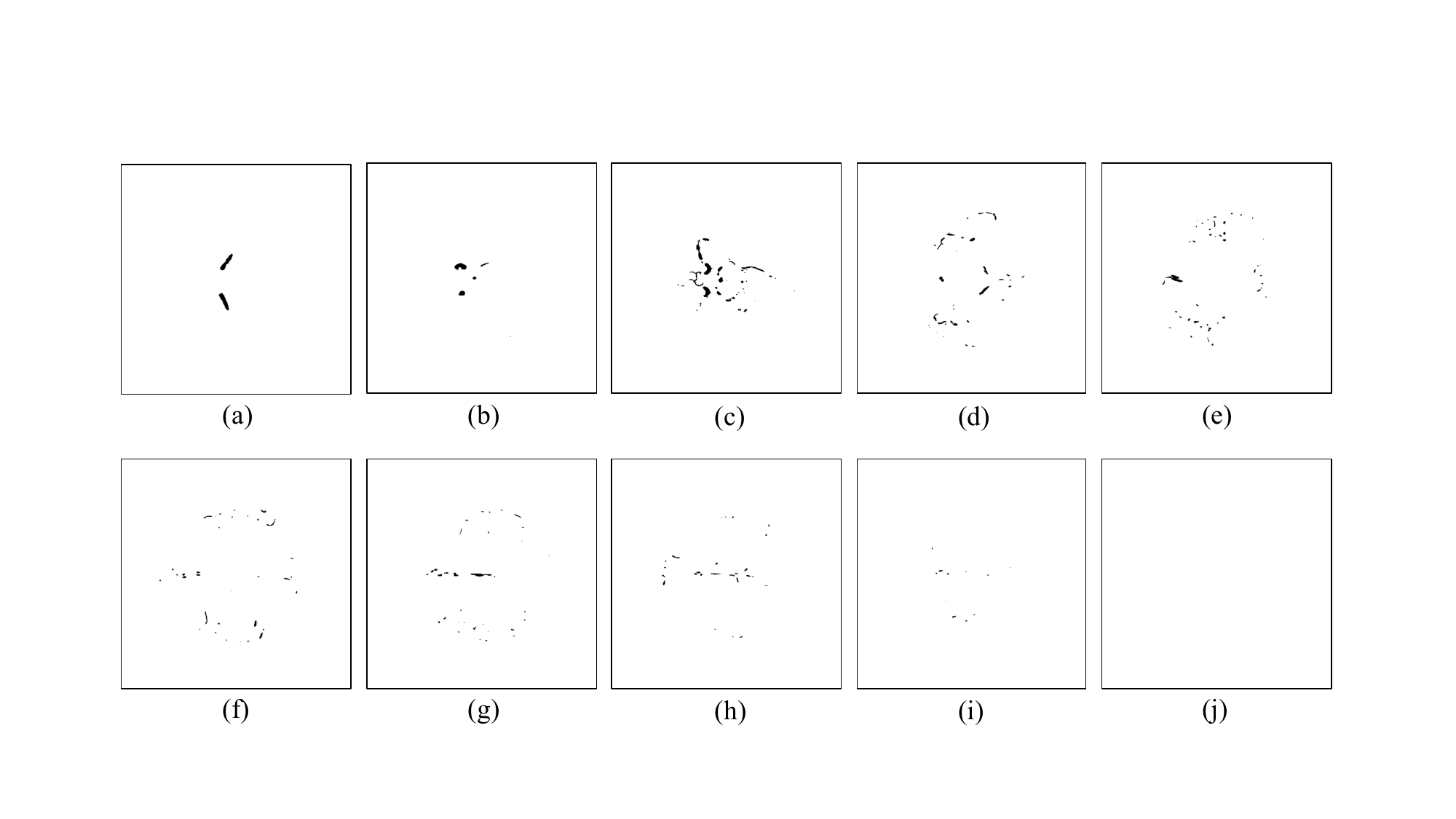}
    \caption{Ten selected slices from the input volume. (a) slice index 1, (b) slice index 15, (c) slice index 29, (d) slice index 43, (e) slice index 57, (f) slice index 71, (g) slice index 85, (h) slice index 99, (i) slice index 113, (j) slice index 128.}
  \label{fig:slices}
\end{figure*}

We use brain vessel point cloud data derived from the validation dataset published in \cite{hilbert2020brave}, which includes labeled images of 20 healthy volunteers across five age groups. The original data comes from a publicly available MRA TOF dataset hosted on the MIDAS data collection website. The matrix and voxel dimensions are $448 \times 448 \times 128$ and $0.51 \times 0.51 \times 0.8$ mm, respectively. The annotations were created by the authors of \cite{hilbert2020brave}, using an initial 2D U-Net pre-segmentation followed by manual correction and cross-verification by junior raters. The dataset is released in NIfTI format (.nii.gz), where each volume includes a binary segmentation mask indicating the presence or absence of blood vessels. Figure~\ref{fig:slices} illustrates several representative slices from the volumetric data.

While the primary dataset for validating our implementation is derived from brain magnetic resonance angiography (MRA), the proposed algorithm is designed for general applicability to any skeleton represented as a 3D point cloud. This includes skeletons from blood vessels, plant roots, connectomes, or other branching tubular networks.

% \subsection{Hardware Information}
% All experiments and prototype development were conducted on a personal laptop equipped with an Apple M1 chip (8-core CPU), 16 GB of unified memory, and running macOS Ventura 13.4 (Build 22F66). The proposed method runs smoothly, processing approximately $60,000$ original point cloud points and performing real-time computations on about $8,000$ skeleton points.
\subsection{Performance}
On typical inputs ($\sim$60k original points, $\sim$8k skeleton points), our prototype supports interactive point selection and returns a correction proposal in real time on a consumer laptop.

\subsection{Pre-processing}
We first load the volumetric labels in Python as voxel grids, then apply 3D skeletonization using the \texttt{skeletonize\_3d} function from the \texttt{skimage.morphology} module. This step extracts the centerlines of segmented vessels, producing a voxel grid where voxels with value $1$ form a point cloud of the vascular tree. This point cloud is then converted into physical coordinates (in \texttt{.xyz} order) and used as input to our algorithm.

\subsection{Algorithm Implementation Platform}
Our algorithm is implemented in C++ using the Libigl library, which offers efficient data structures for managing vertices and edges, along with tools for user interaction. We use the built-in viewer to render the point cloud and allow point selection via mouse clicks. A graphical menu enables users to invoke algorithmic functions such as propagation and correction. The code is available at \url{https://github.com/yangruoxuan0516/ParcoursRecherche}.

\subsection{User Interaction}
\begin{figure}[htbp!]
  \centering
  \includegraphics[width=1\linewidth]{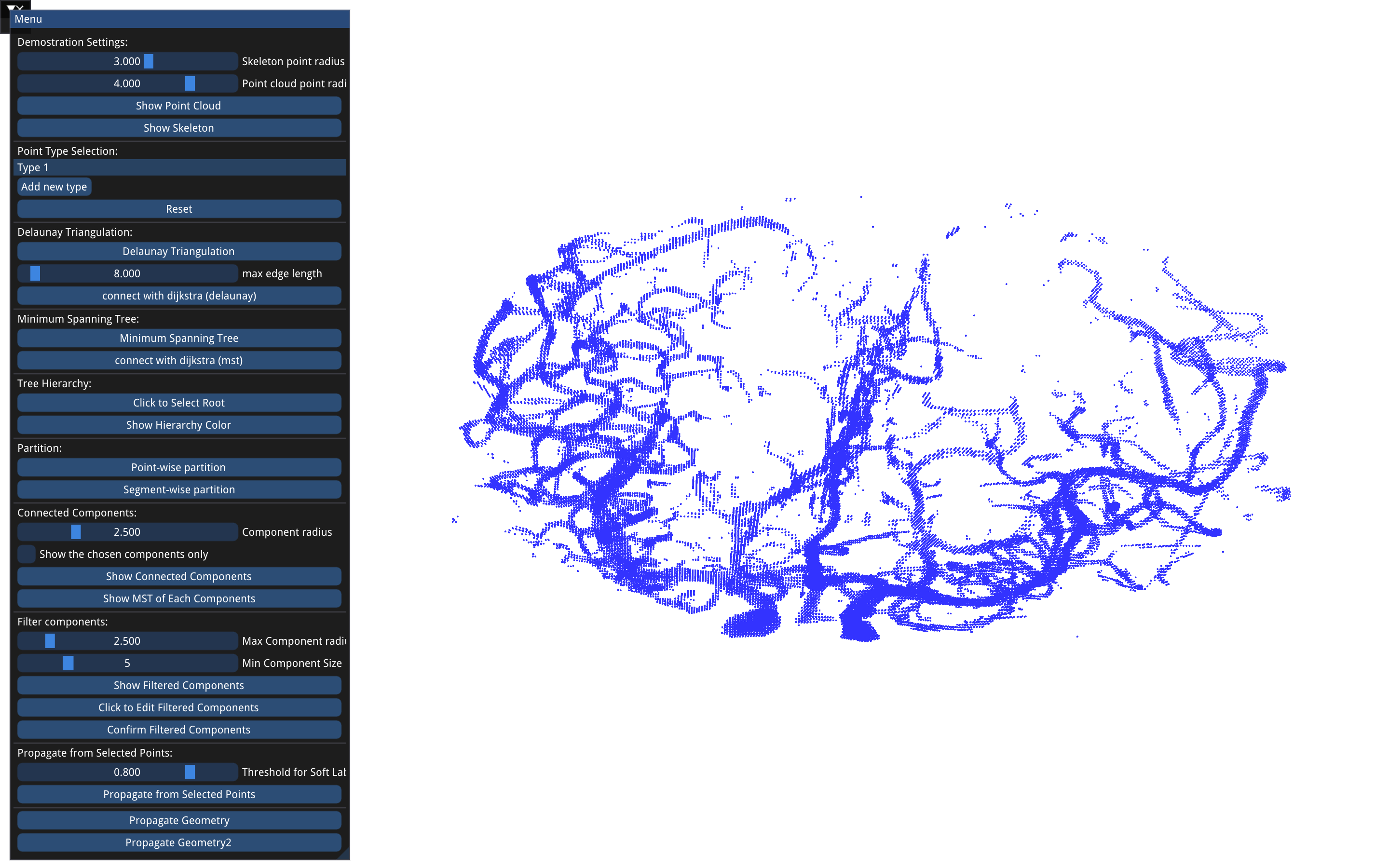}
  \caption{The user interface of our proposed algorithm, implemented with Libigl in C++. The menu appears on the left by default, and the point cloud is rendered in the center.}
  \label{fig:ui}
\end{figure}

Using the Libigl viewer, as shown in Figure~\ref{fig:ui}, we display both the original vessel point cloud and the corresponding skeleton. Users can freely navigate the 3D view, select source and target points for correction, and initiate path propagation. In the case of a “dotted vessel” or a “crossing vessel,” as illustrated in Figure~\ref{fig:teaser}, users can select a single source-target pair or multiple such pairs. The algorithm is then triggered by selecting the appropriate function from the menu.

\subsection{Main Algorithm}
The main algorithm takes the extracted centerline point cloud and a set of $2n$ user-specified points as input, and automatically constructs $n$ plausible vessel paths. Each $i^{th}$ path uses the $2i^{th}$ and $2i+1^{th}$ points as its source and target, respectively.
% \paragraph{Graph definitions.}
\medskip
\noindent\textbf{Graph definitions.}
Let $V=\{v_i\}$ be the set of 3D skeleton points in physical coordinates. We use two auxiliary graphs:
(1) a \emph{distance-threshold} neighborhood graph $G_{\delta}=(V,E_{\delta})$ where $(v_i,v_j)\in E_{\delta}$ iff $\|v_i-v_j\|<\delta$, used only to extract connected components via BFS;
(2) a \emph{3D Delaunay edge graph} $G_{\mathrm{dt}}=(V,E_{\mathrm{dt}})$, where $E_{\mathrm{dt}}$ is the 1-skeleton (edge set) of the Delaunay tetrahedralization of $V$.
We further define the filtered Delaunay graph $G_{\mathrm{dt}}^{\tau}=(V,E_{\mathrm{dt}}^{\tau})$ with $E_{\mathrm{dt}}^{\tau}=\{e\in E_{\mathrm{dt}}:\|e\|<\tau\}$.
For each connected component of $G_{\delta}$, we compute a minimum spanning tree (MST) using Euclidean edge weights; the union over components forms our component-wise MST graph $G_{\mathrm{mst}}$.

\subsubsection{Overview}
\begin{figure}[htbp!]
  \centering
  \includegraphics[width=\linewidth, trim=100 0 100 0, clip]{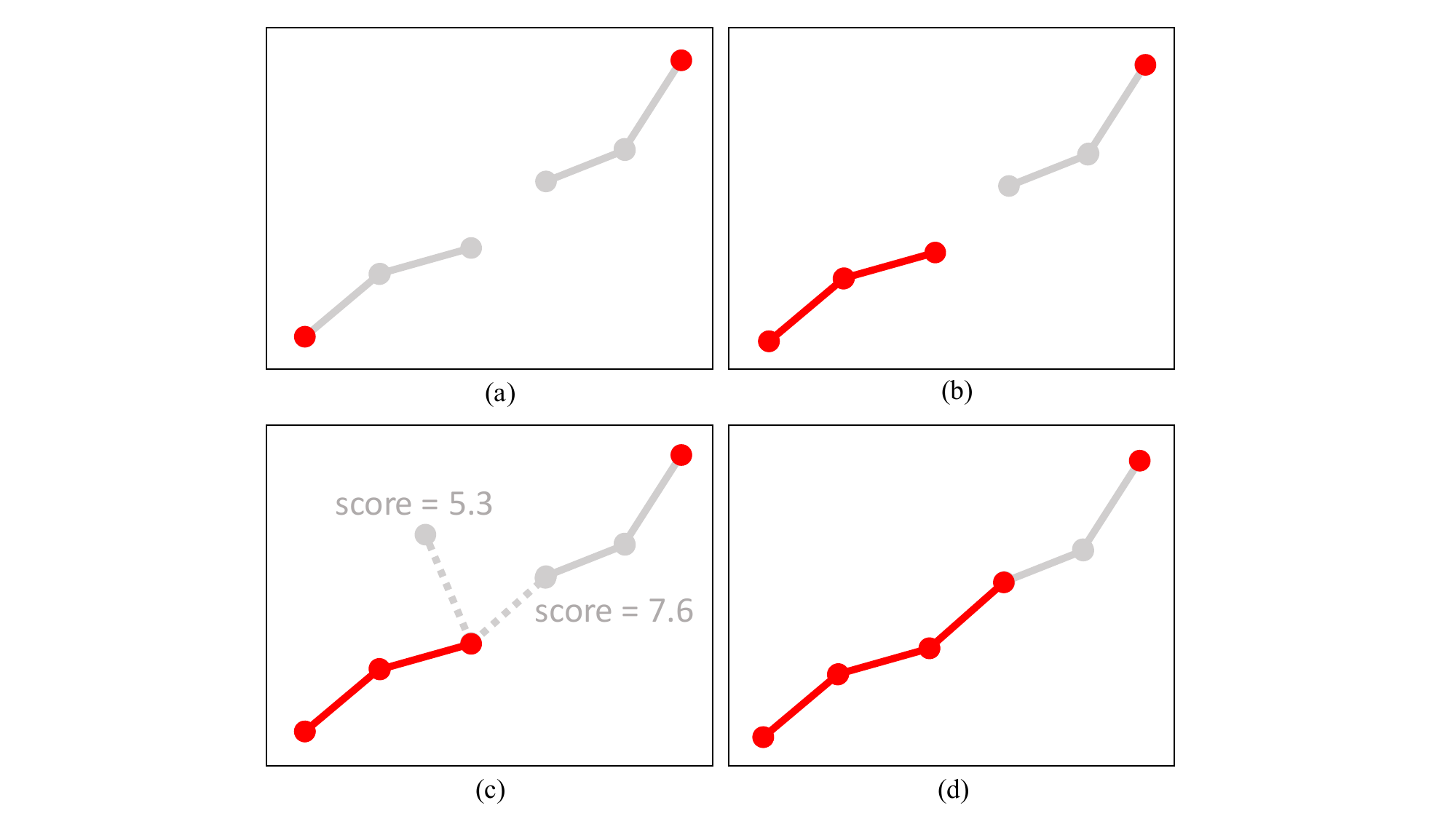}
  \caption{An overview of our semi-automatic vessel tracing process. (a) The source-target pair is selected. (b) Propagation is performed along edges of the component-wise minimal spanning tree. (c) When an endpoint or branching point is encountered, all Delaunay edges below a length threshold are evaluated. (d) The best-scoring edge is selected as the next propagation step.}
  \label{fig:demo}
\end{figure}

The full procedure of our semi-automatic vessel tracing method for each source-target pair is shown in Figure~\ref{fig:demo} and described in Algorithm \ref{alg:vessel_pathfinding}. It can be applied to both “crossing vessels” and “dotted vessels.”

\begin{algorithm}
\caption{Semi-Automatic Vessel Tracing}
\label{alg:vessel_pathfinding}
\footnotesize
\begin{algorithmic}[1]
\Require Source $s$, Target $t$, Skeleton point set $V$, Component-wise MST edges $E_{\text{mst}}$, Filtered Delaunay edges $E_{\text{dt}}$
\Ensure Reconstructed path from $s$ to $t$

\State Initialize $P \gets [s]$, $visited \gets \{s\}$
\State Construct MST adjacency list $\mathcal{A}_{\text{mst}}$
\State Construct Delaunay adjacency list $\mathcal{A}_{\text{dt}}$
\State $v \gets s$

\If{$v$ has one unvisited neighbor in $\mathcal{A}_{\text{mst}}$}
    \State Propagate along MST until a junction or endpoint is reached
    \State Append visited nodes to $P$
\ElsIf{$v$ has two unvisited neighbors}
    \State Propagate along both directions temporarily
    \State Estimate distance to $t$ via Dijkstra
    \State Choose direction with shorter estimated distance
    \State Continue propagation from selected direction
\EndIf

\While{$v \neq t$ \textbf{and} max iterations not exceeded}
    \State $N_{\text{mst}} \gets$ unvisited MST neighbors of $v$
    
    \If{$|N_{\text{mst}}| = 1$}
        \State $v \gets N_{\text{mst}}[0]$
        \State Append $v$ to $P$, mark as visited
    \Else
        \State $N_{\text{dt}} \gets$ unvisited Delaunay neighbors of $v$
        \If{$N_{\text{dt}} = \emptyset$}
            \State \textbf{break} \Comment{No path found}
        \EndIf

        \State Initialize best score $S^\ast \gets -\infty$, best candidate $v^\ast \gets \varnothing$
        \ForAll{$u \in N_{\text{dt}}$}
            \State $S \gets 0$

            \If{$u = t$} \State $S \gets S + w_{\text{target}}$ \EndIf
            \If{$u$ and $v$ in same component} \State $S \gets S + w_{\text{component}}$ \EndIf

            \State $\vec{d}_{\text{prev}} \gets$ direction of last $n$ steps
            \State $\vec{d}_{vu} \gets$ unit vector from $v$ to $u$
            \State $\theta \gets \arccos(\vec{d}_{\text{prev}} \cdot \vec{d}_{vu})$
            \State $S \gets S - w_{\text{angle}} \cdot \theta$

            \State $\ell \gets \|V_u - V_v\|$
            \State $S \gets S - w_{\text{length}} \cdot \ell$

            \State $d_t \gets$ Dijkstra distance from $u$ to $t$
            \State $S \gets S - w_{\text{distance}} \cdot d_t$

            \If{$S > S^\ast$}
                \State $S^\ast \gets S$, $v^\ast \gets u$
            \EndIf
        \EndFor

        \If{$v^\ast = \varnothing$}
            \State \textbf{break} \Comment{No viable candidate}
        \Else
            \State $v \gets v^\ast$
            \State Append $v$ to $P$, mark as visited
        \EndIf
    \EndIf
\EndWhile

\State \Return $P$
\end{algorithmic}
\end{algorithm}

\subsubsection{Component-wise MST}
\label{sec:component}
As described in the pseudocode, when tracing the vessel centerline, if the current point has only a single unvisited neighbor in the MST graph, we proceed to that neighbor. Since our method uses geometric criteria to determine the next step, we prefer to follow the MST whenever possible. This is because MSTs have simpler structures compared to other graphs, such as Delaunay triangulations, and are less prone to error—especially when the skeleton is noisy or non-smooth. Furthermore, following the MST facilitates efficient computation.

Rather than using a single MST over the entire skeleton point cloud, we first divide the point cloud into several components based on pairwise distance: two points belong to the same component only if their distance is below a threshold (heuristically set to $2.5$). This is achieved by constructing a distance-aware adjacency list and applying BFS to extract connected components. The component-wise MSTs inherit the robustness of traditional MSTs while excluding excessively long edges that are unlikely to belong to valid vessel paths, thereby reducing potential errors in subsequent steps.
\begin{figure}[htbp!]
  \centering
  \includegraphics[width=0.9\linewidth, trim=220 60 220 80, clip]{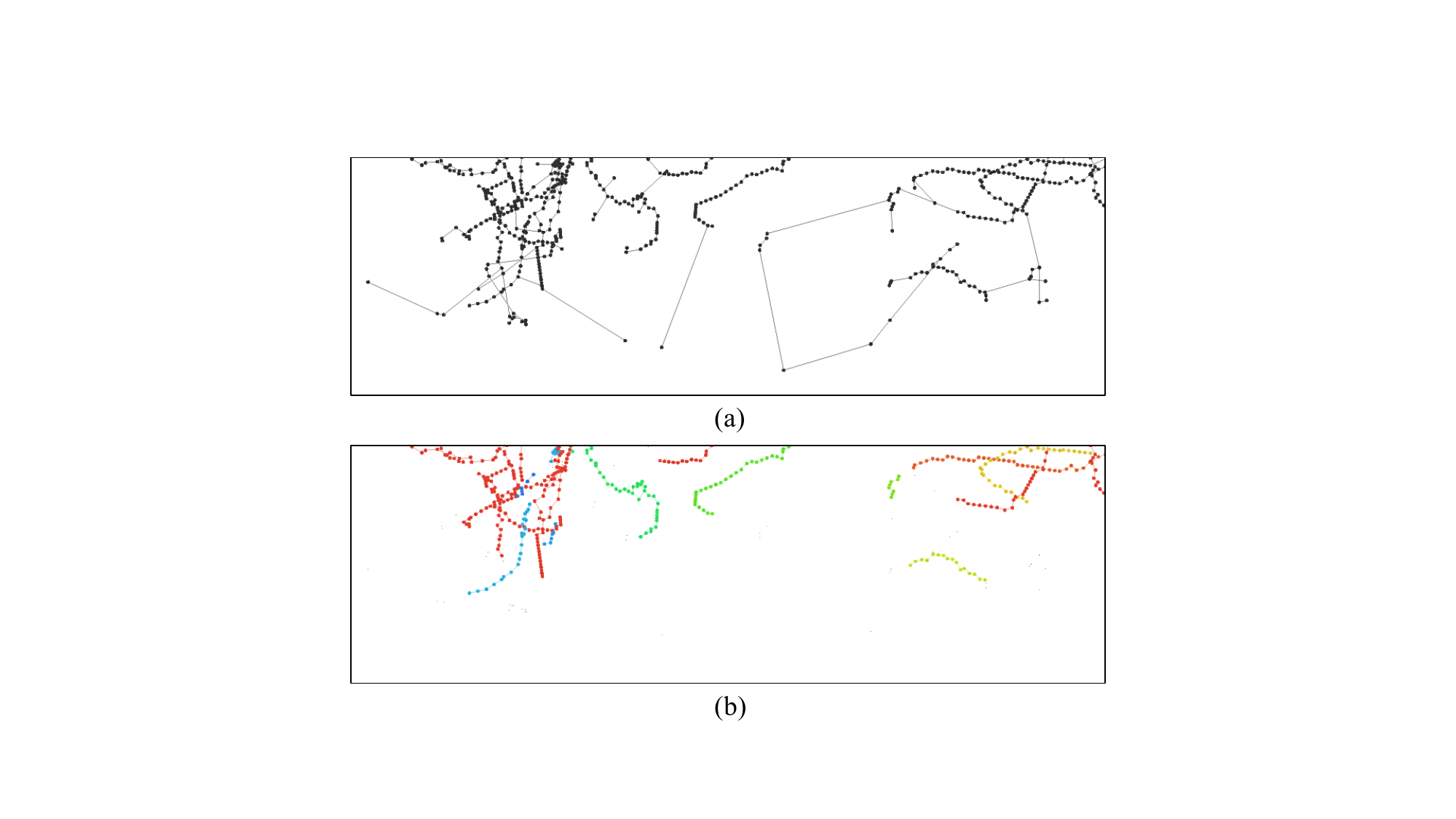}
  \caption{(a) The minimum spanning tree of the entire point cloud may introduce incorrect connections due to isolated distant points. (b) Component-wise minimum spanning trees ignore overly long, unnatural connections; different components are indicated in different colors.}
  \label{fig:componentMST}
\end{figure}

\subsubsection{Neighbor Selection in Delaunay Triangulation}
\label{sec: Neighbor choosing in Delaunay triangulation}

If MST tracing is not possible—i.e., when an endpoint (no unvisited neighbors) or a branching point (multiple unvisited neighbors) is reached in the component-wise MST—we switch to using filtered Delaunay triangulation edges. Specifically, we consider only those edges shorter than a threshold (heuristically set to $8.0$). Each unvisited neighbor is scored using the following function:
% \begin{align*}
%     \text{score} =\ & w_{\text{target}} \cdot \mathcal{T} \\
%     & + w_{\text{component}} \cdot \mathcal{C} \\
%     & - w_{\text{angle}} \cdot \mathcal{A} \\
%     & - w_{\text{length}} \cdot \mathcal{L} \\
%     & - w_{\text{distance}} \cdot \mathcal{D}
% \end{align*}

\begin{equation}
\label{eq:score}
\begin{aligned}
S(u\mid v) =\;& w_{\mathrm{tar}}\,\mathcal{T}(u) + w_{\mathrm{comp}}\,\mathcal{C}(u,v) \\
&- w_{\mathrm{ang}}\,\mathcal{A}(u,v) - w_{\mathrm{len}}\,\mathcal{L}(u,v) - w_{\mathrm{dist}}\,\mathcal{D}(u,t).
\end{aligned}
\end{equation}

We choose the next step by $u^\star = \arg\max_{u \in \mathcal{N}_{\mathrm{dt}}(v)} S(u \mid v)$.
Each term in the score function is defined as follows:
\begin{itemize}
    \item $\mathcal{T} = \mathbf{1}_{\{u = t\}}$ (Target bonus): Indicator function that equals 1 if the candidate point $u$ is the target $t$, and 0 otherwise. A large positive weight encourages the algorithm to select the target point immediately if it is reachable.

    \item $\mathcal{C} = \mathbf{1}_{\{u \sim v\}}$ (Component bonus): Equals 1 if the candidate $u$ and current point $v$ belong to the same connected component, and 0 otherwise. This prevents premature jumps across components.

    \item $\mathcal{A} = \arccos\left( \frac{\vec{d}_{\text{prev}} \cdot \vec{d}_{vu}}{\|\vec{d}_{\text{prev}}\| \cdot \|\vec{d}_{vu}\|} \right)$ (Angle penalty): Penalizes sharp direction changes. The vector $\vec{d}_{\text{prev}}$ is estimated from the last $n$ points in the current path ($n=4$ heuristically), while $\vec{d}_{vu}$ is the direction vector from $v$ to $u$.

    \item $\mathcal{L} = \|V_u - V_v\|$ (Length penalty): Euclidean distance between the current point $v$ and the candidate $u$. Shorter steps are preferred to maintain local coherence.

    \item $\mathcal{D} = \text{Dijkstra}(u, t)$ (Distance penalty): Estimated shortest path length from $u$ to the target $t$, computed over the filtered Delaunay graph. This encourages global progression toward the goal.
\end{itemize}
Each factor is weighted by a corresponding coefficient: $w_{\text{target}}$, $w_{\text{component}}$, $w_{\text{angle}}$, $w_{\text{length}}$, and $w_{\text{distance}}$. The values used in our implementation are: $1000$, $5$, $10$, $1$, and $\frac{1}{2}$, respectively, chosen heuristically to balance accuracy and smoothness.

\subsubsection{Start Direction Selection}
As mentioned in Section \ref{sec: Neighbor choosing in Delaunay triangulation}, the path direction is computed based on the last $n$ points. However, when only the source point is provided at the beginning, the initial direction is ambiguous—even in the MST, where only two directions are possible. Therefore, we initiate propagation in both directions from the source point until reaching a terminal or branching point (or the target itself). We then compare the Dijkstra distances in the Delaunay triangulation from the two resulting endpoints to the target. The direction with the shorter distance is selected for further propagation, while the other endpoint is treated as the terminal end of a vessel.

\subsection{Parameter Selection}
\label{sec:parameter_choosing}
As described in the algorithm, numerous parameters are chosen heuristically, and their values significantly influence the propagation results. It would be interesting to utilize annotated data and apply optimization techniques to determine the best parameter set, or to adapt parameters dynamically based on local geometric conditions. However, in this work, we assign fixed values based on empirical observations and data characteristics.

\subsubsection{Delaunay Filter Distance}
\begin{figure}[htbp!]
  \centering
  \includegraphics[width=\linewidth, trim=120 0 230 0, clip]{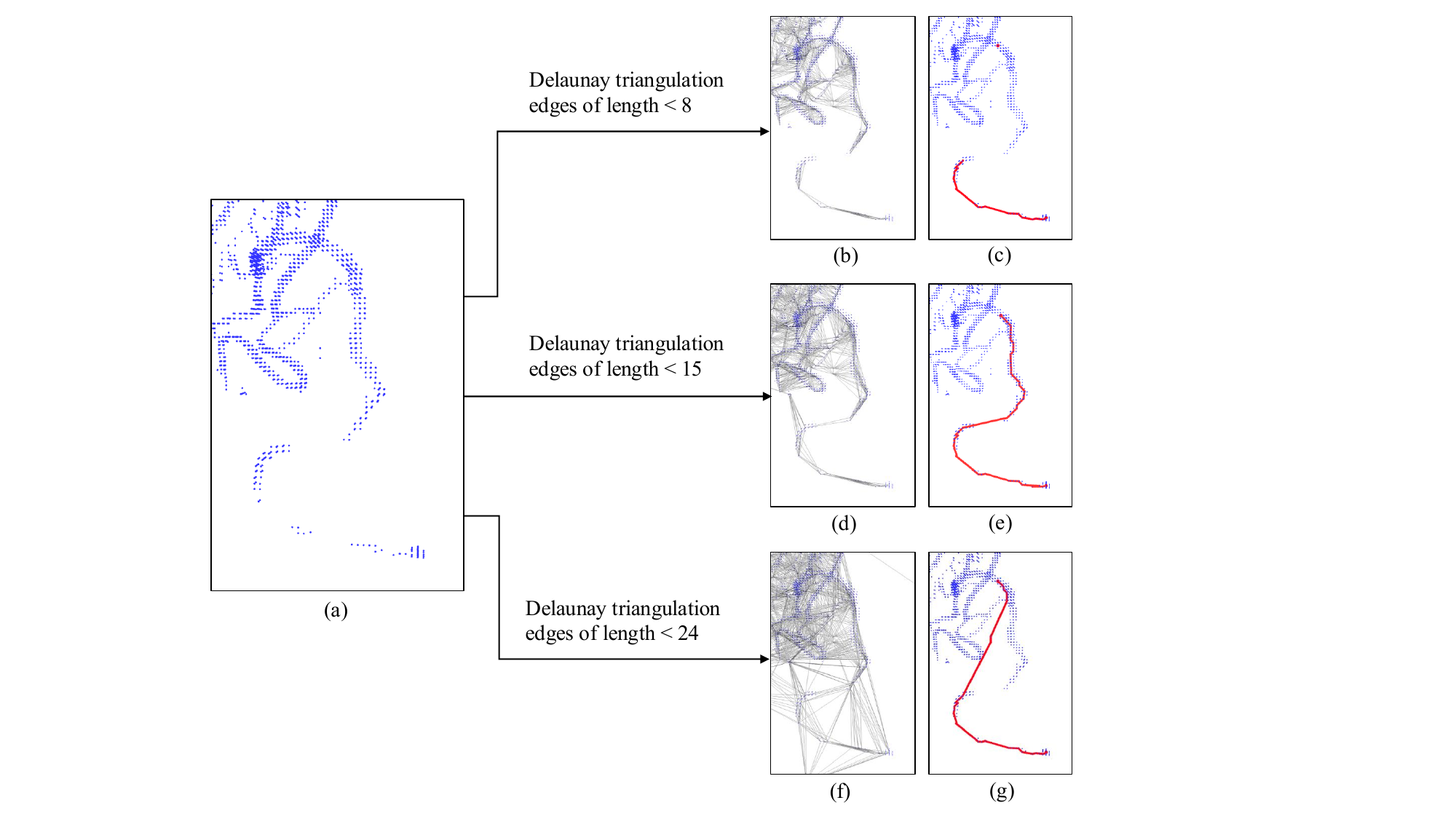}
  \caption{(a) The original point cloud. (b) Delaunay triangulation edges with length $< 8$. (c) Propagation fails due to insufficient connectivity. (d) Edges with length $< 15$. (e) Propagation succeeds with proper connection. (f) Edges with length $< 24$. (g) Propagation selects an incorrect path due to a smaller turning angle.}
  \label{fig:dt_param}
\end{figure}

During propagation, when the current point is either a terminal or a branching node and the MST path cannot be followed, we consider neighbors in the filtered Delaunay graph as candidates for the next step. The choice of distance threshold determines the candidate set. If the threshold is too small, the true next node may be disconnected from the filtered graph and thus excluded from consideration (see Figure~\ref{fig:dt_param}). On the other hand, a threshold that is too large may introduce incorrect candidates, which can both outscore the correct one and slow down score computation. Unless otherwise stated, all distance thresholds are applied in the same coordinate system as the skeleton point set (physical units after voxel-to-world conversion).

We choose the threshold to ensure that each component (as defined in Section \ref{sec:component}) remains internally connected.

\subsubsection{Number of Previous Points for Direction Estimation}
Due to noise in the skeletonization results, we estimate the propagation direction using the last $n$ points, rather than just the previous one. This strategy mitigates spurious twists and irregularities. If $n$ is too small, it may fail to smooth out noise; if $n$ is too large, the estimated direction may become outdated and unresponsive to recent changes.

The current value of $n$ is fixed, but an adaptive strategy could be beneficial—for example, reducing $n$ in regions of high curvature to reflect rapid directional changes, and increasing it in straighter segments to suppress noise.
% TODO：figure n太小没变化 太大就错过变化

\subsubsection{Score Weights}
The scoring function combines multiple criteria (angle, step length, target distance, etc.), each with its own scale—for instance, angles range within $[0, \pi]$, whereas point distances are typically close to $1$. Therefore, weights are introduced both to normalize units and to reflect the relative importance of different terms. For example, angle penalties are given more weight than step length penalties, as abrupt directional changes are generally less desirable than small deviations in distance.

Exploring different parameter combinations or formulating a learning-based optimization to tune them would be promising future work. However, due to time constraints and the lack of annotated data, we leave this for future exploration.

\section{Experiments \& Evaluation}
\begin{figure}[htbp!]
  \centering
  \includegraphics[width=\linewidth, trim=300 25 300 20, clip]{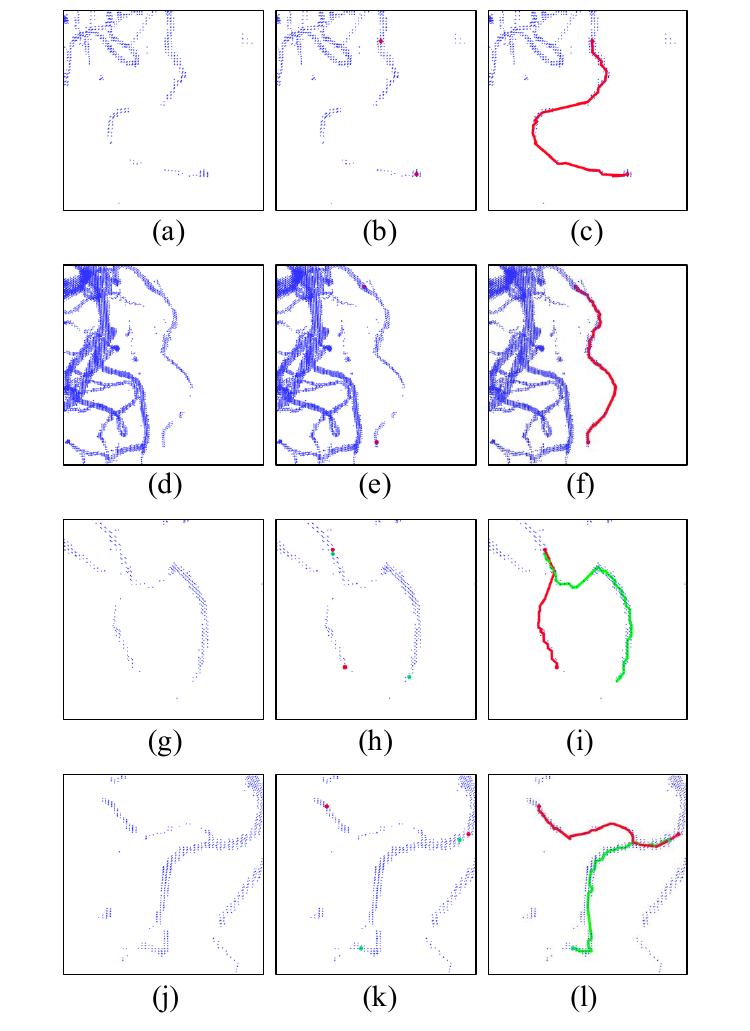}
  \caption{(a), (d), (g), (j): Original point cloud. (b), (e), (h), (k): Selected source-target pairs. (c), (f), (i), (l): Resulting connections.}
  \label{fig:examples}
\end{figure}

We evaluate our method on multiple cases of “crossing vessels” and “dotted vessels” identified within the brain vessel dataset. Representative examples are shown in Figure~\ref{fig:examples}. The left column displays the original point cloud, the middle column shows the filtered Delaunay triangulation edges, and the right column presents the propagation results. We use a default Delaunay filtering threshold $\tau$ and only increase it when the filtered graph becomes disconnected around the user-specified endpoints. In practice, selecting the smallest $\tau$ that preserves intra-component connectivity provides a reproducible and stable candidate neighborhood, while avoiding overly long edges that introduce clutter and ambiguous shortcuts.

% As demonstrated, our method successfully reconstructs anatomically plausible connections in the majority of test cases. Failure cases are discussed in \ref{sec:fail}
As demonstrated in representative cases, our method often reconstructs plausible connections. We further discuss typical failure modes in \ref{sec:fail}.

While the current evaluation focuses on a specific class of tubular structures, the underlying approach is inherently general and designed to support a wide range of applications. Future work will include quantitative assessments on diverse datasets—such as plant root architectures and skeletons extracted from connectomic reconstructions—where analogous topological artifacts frequently occur and demand efficient correction strategies.

\section{Qualitative Comparison to Prior Interactive Correction} \label{sec:compare_with_invaner}

\begin{figure}[htbp!]
  \centering
  \includegraphics[width=\linewidth]{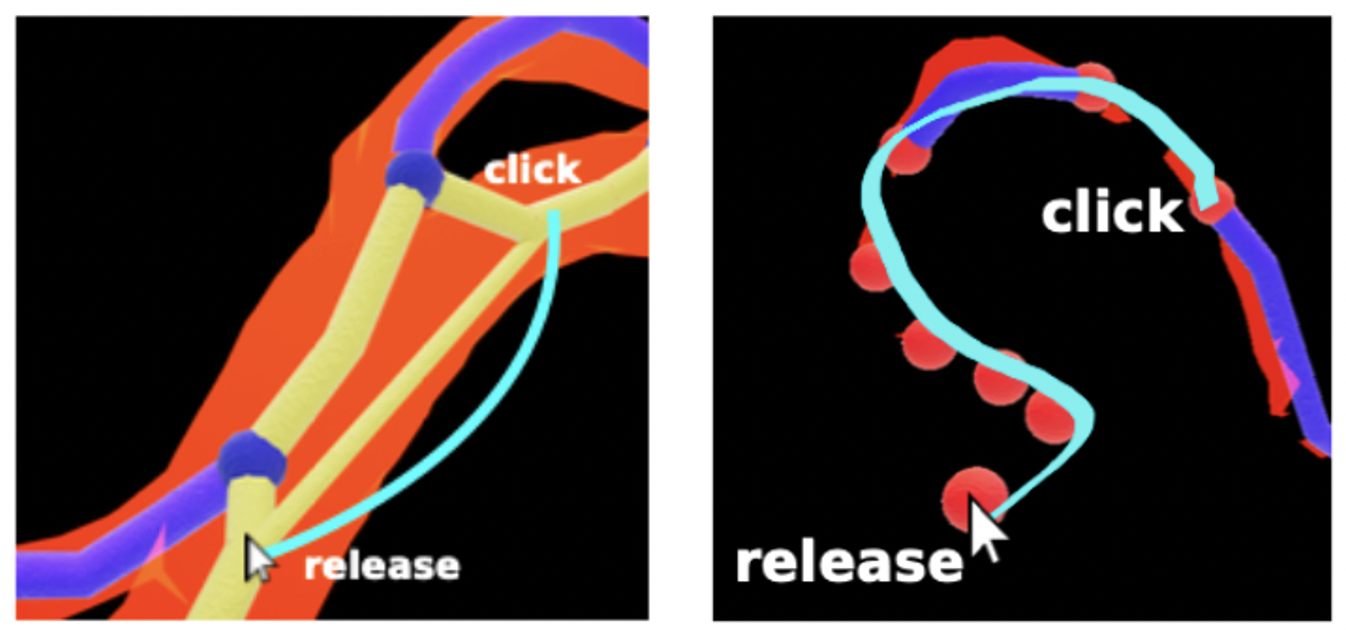}
  \caption{Illustration of INVANER tool usage. Left: separating "kissing vessels" by dragging across the shared segment. Right: reconnecting "dotted vessels" by dragging over all disconnected segments. \cite{nigolian2019invaner}}
  \label{fig:INVANER}
\end{figure}

The proposed method is semi-automatic and, compared to the manual correction method introduced in \cite{nigolian2019invaner}, aims to reduce interaction effort while providing a comparable correction workflow in the two targeted failure modes. In \cite{nigolian2019invaner}, to address the “kissing vessel” problem—where vessel 1 and vessel 2 erroneously share a common branch due to upstream errors, as illustrated in Figure~\ref{fig:INVANER} (left)—the user must click on one end of vessel 1, drag the cursor to the other end, and release in order to define a new path for vessel 1 and thereby separate the overlapping vessels.

To fix a “dotted vessel,” as shown in Figure~\ref{fig:INVANER} (right), the user is required to click on the first segment, drag the cursor across all disconnected fragments, and release on the final segment to validate the connection.

Although these interactions seem intuitive and flexible, they present some limitations. First, they can be tedious and error-prone, particularly in the “dotted vessel” case, where missing a single segment during dragging can invalidate the entire operation, requiring the user to repeat the process. Second, the results are highly dependent on the user’s precision, which may lead to deviations from the original anatomical geometry—e.g., in the “kissing vessel” case, if the user fails to click and release near the shared segment, the newly created path may diverge from the intended vessel.

In contrast, our proposed semi-automatic method combines flexibility with geometric fidelity. While users can still intervene, the automatic propagation process reduces manual workload and ensures consistency with the underlying data. For both “crossing segments” (where two separate segments intersect and form a topological loop—similar in spirit to “kissing vessels”) and “dotted segments,” the user only needs to select a source and target point at either end of the desired connection. The algorithm then computes a plausible path that either separates the loop (in the case of "crossing segments") or reconnects the disjoint segments (in the case of "dotted segments").

% Importantly, the algorithm operates entirely on the geometric features of the skeleton point cloud and does not modify the data during correction, thereby preserving the consistency and ensuring that the output remains faithful to the original scan.

Importantly, the algorithm operates on the geometric features of the skeleton point cloud and outputs an ordered polyline (edge sequence) that can be accepted by the user and integrated into downstream skeleton editing workflows.

\begin{figure}[htbp!]
  \centering
  \includegraphics[width=\linewidth, trim=180 250 170 130, clip]{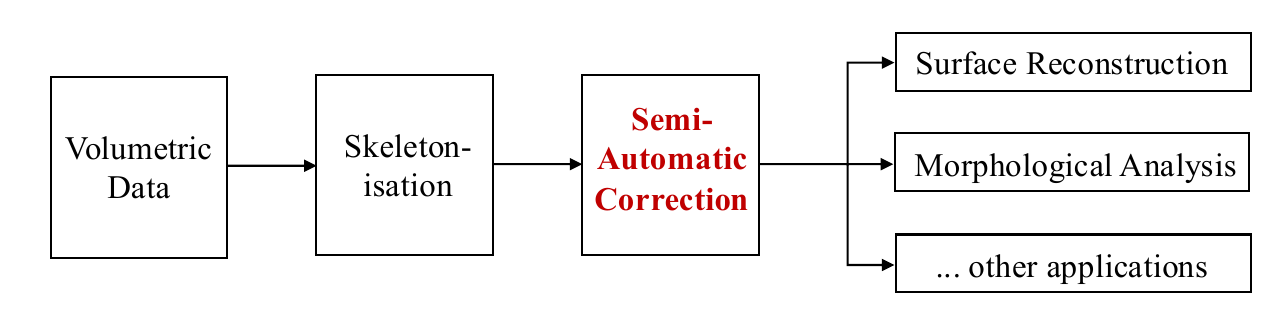}
  \caption{The complete workflow, consisting of automatically generated results followed by manual corrections augmented with semi-automatic methods.}
  \label{fig:workflow}
\end{figure}

\section{Observations \& Future Works}
\label{sec:fail}
Due to time constraints, the proposed method still possesses several limitations. First of all, although it theoretically requires less user interaction and demonstrates greater efficiency, a dedicated user study is necessary to evaluate its real-world performance. Furthermore, annotated datasets from various domains—including clinical vascular imaging, agricultural root scans, and large-scale connectomic reconstructions—must be acquired to quantitatively assess the algorithm’s accuracy and robustness across diverse tubular structures.

Second, the current implementation remains a prototype and is not yet integrated into a complete workflow. Ideally, we hope to incorporate it as a post-processing step after the automatic skeletonization, assisting manual correction as illustrated in the overall processing pipeline shown in Figure~\ref{fig:workflow}.

\begin{figure}[htbp!]
  \centering
  \includegraphics[width=\linewidth, trim=60 50 60 60, clip]{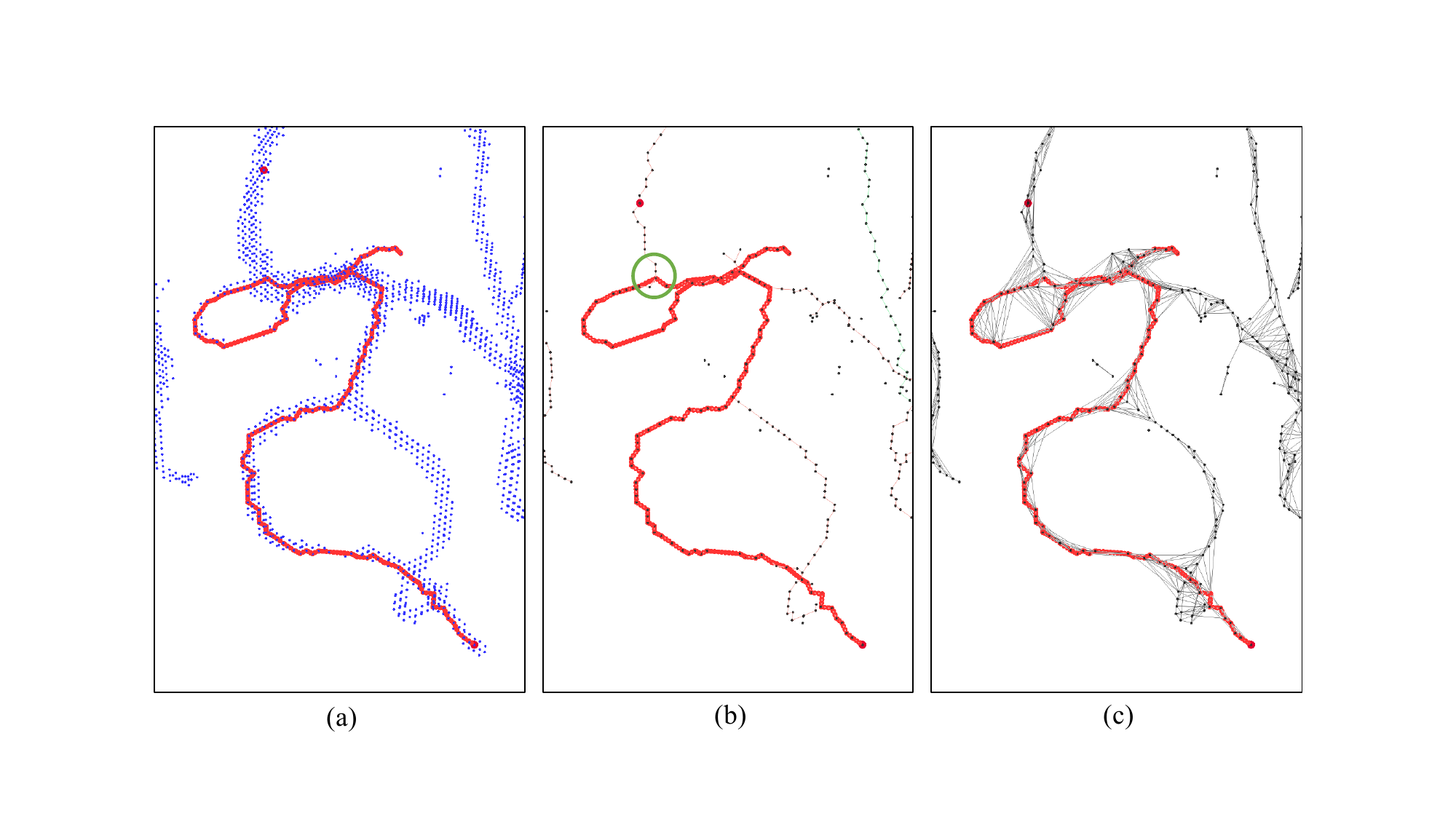}
  \caption{(a) The selected source point is at the lower end of the red path. The target point, located in the upper part of the image, is not reached. (b) The green circle highlights the key decision point where propagation goes in the wrong direction due to a smaller angle. (c) Delaunay triangulation edges are cluttered near the error region.}
  \label{fig:wrong}
\end{figure}

Third, in certain edge cases—such as when the topological structure between the chosen source and target points is overly complex, as illustrated in Figure~\ref{fig:wrong}—the algorithm may fail and generate incorrect results. A more systematic collection of failure cases is needed for comprehensive evaluation. Furthermore, improvements to the propagation criteria and implementation of robust error-handling mechanisms remain necessary.

In this particular example, the path is mistakenly guided to the left instead of the intended right direction at the decision point marked by the green circle. This misdirection primarily occurs because the turning angle toward the incorrect path is smaller.

As a result, several aspects remain open for future exploration. First, as discussed in Section~\ref{sec:parameter_choosing}, the parameters used in the algorithm could be more rigorously optimized—either by formulating a formal optimization problem with annotated data or by dynamically adapting parameter values based on local data characteristics.

Second, annotated datasets must be acquired to quantitatively assess the algorithm’s accuracy. Moreover, a well-designed user study is essential to evaluate the efficiency of the method and its effectiveness in facilitating skeleton correction.

Finally, if the approach proves robust and reliable, it would be valuable to integrate it into a comprehensive skeleton correction workflow. For instance, as in the connectome correction system proposed by Dorkenwald et al. \cite{dorkenwald2023cave}, our method could augment large-scale interactive annotation tools and enhance user experience.

% \subsection{Original point cloud and skeleton point cloud} \begin{figure}[htbp!] \centering \includegraphics[width=\linewidth, trim=40 125 50 120, clip]{fig/point_cloud.pdf} 
% left bottom right top \caption{(a) The original point cloud. (b) The skeleton point cloud.} \label{fig:point_cloud} 
% \end{figure} 

\section{Conclusion}

% \subsection{Component and minimal spanning tree (MST)} \begin{figure}[htbp!] \centering \includegraphics[width=1.1\linewidth, trim=160 10 160 20, clip]{fig/mst.pdf} 
% left bottom right top \caption{(a) The MST of all skeleton point cloud. (b) The skeleton point cloud is separated into different components, with different components indicated in different colors. (c) The component-wise MST (refer to Section~\ref{sec:component}) for the definition of component.} \label{fig:mst} 
% \end{figure} 

% \subsection{Filtered Delaunay triangulation} 

% \begin{figure}[htbp!] \centering \includegraphics[width=1.1\linewidth, trim=50 10 50 10, clip]{fig/dt.pdf} left bottom right top \caption{(a) The Delaunay triangulation edges with length $< 4$. (b) Edges with length $<8$. (c) Edges with length $<12$. (d) Edges with length $<16$.} \label{fig:dt} \end{figure}

\begin{figure}[htbp!]
  \centering
  \includegraphics[width=1.1\linewidth, trim=50 10 50 10, clip]{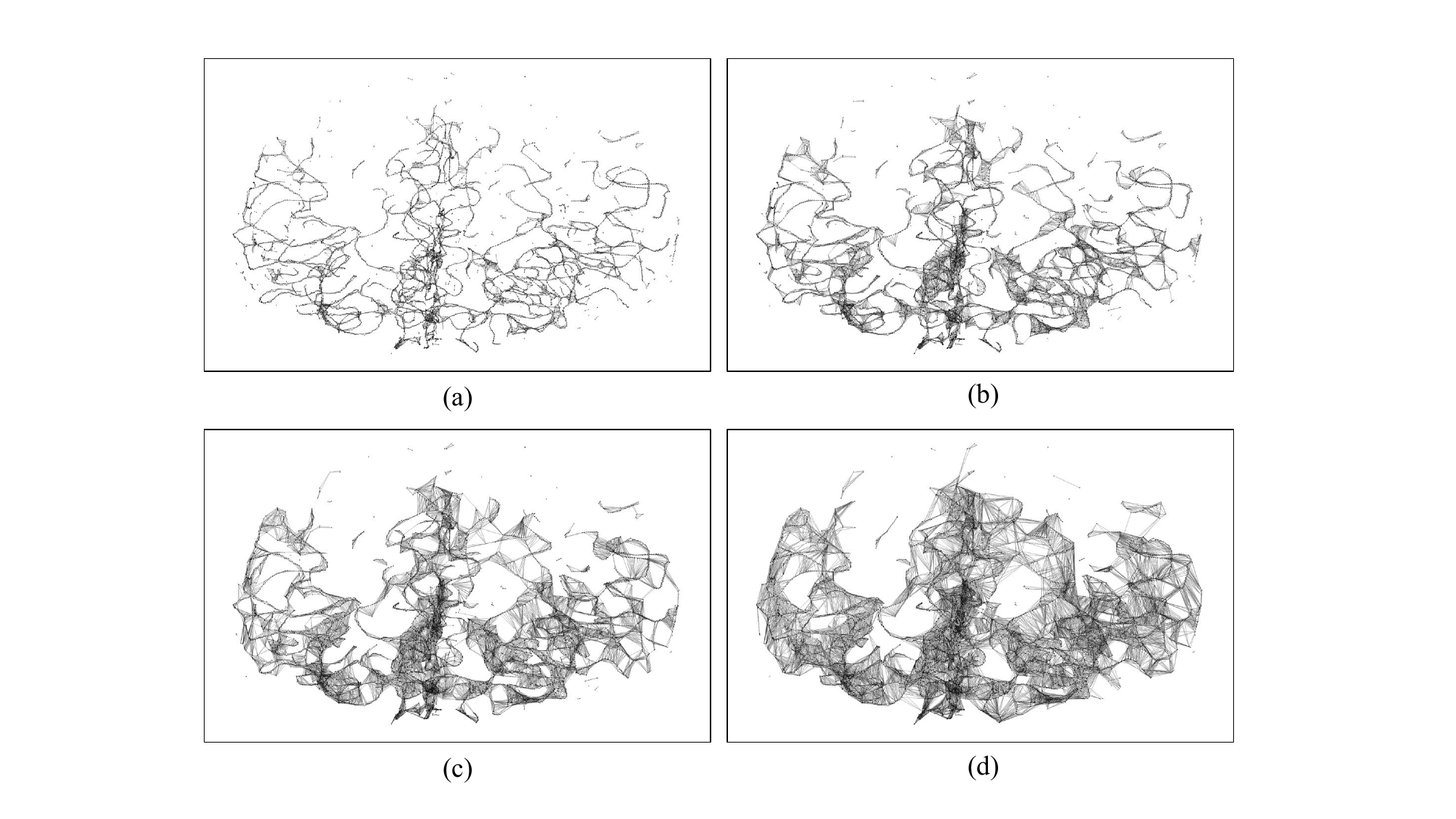}
  \caption{(a) The Delaunay triangulation edges with length $< 4$. (b) Edges with length $< 8$. (c) Edges with length $< 12$. (d) Edges with length $< 16$.}
  \label{fig:dt}
\end{figure}
We proposed a semi-automatic correction algorithm for skeletons of 3D tubular structures that balances minimal user interaction with geometric and topological reasoning. By combining component-wise MST-based propagation with filtered Delaunay triangulation heuristics, our method effectively reconstructs morphologically plausible paths from sparse user inputs. Qualitative evaluations on brain vessel datasets demonstrate the framework’s ability to resolve common skeletonization artifacts such as “crossing segments” and “dotted segments”. The underlying principles suggest applicability to other complex networks like vascular structures, root systems and neural connectomes. Although the current implementation lacks full workflow integration and large-scale quantitative validation across these diverse domains, it provides a robust and efficient foundation for future interactive analysis tools in biomedical imaging, plant sciences, and neuroinformatics, facilitating more accurate and efficient research on various tubular systems.

\bibliography{sn-bibliography}% common bib file

@article{hilbert2020brave,
  title={BRAVE-NET: fully automated arterial brain vessel segmentation in patients with cerebrovascular disease},
  author={Hilbert, Adam and Madai, Vince I and Akay, Ela M and Aydin, Orhun U and Behland, Jonas and Sobesky, Jan and Galinovic, Ivana and Khalil, Ahmed A and Taha, Abdel A and Wuerfel, Jens and others},
  journal={Frontiers in artificial intelligence},
  volume={3},
  pages={552258},
  year={2020},
  publisher={Frontiers Media SA}
}

@inproceedings{nigolian2019invaner,
  title={INVANER: INteractive vascular network editing and repair},
  author={Nigolian, Valentin Z and Igarashi, Takeo and Seo, Hirofumi},
  booktitle={Proceedings of the 32nd annual ACM symposium on user interface software and technology},
  pages={1197--1209},
  year={2019}
}

@article{kirbas2004review,
  title={A review of vessel extraction techniques and algorithms},
  author={Kirbas, Cemil and Quek, Francis},
  journal={ACM Computing Surveys (CSUR)},
  volume={36},
  number={2},
  pages={81--121},
  year={2004},
  publisher={ACM New York, NY, USA}
}

@article{selvan2016extraction,
  title={Extraction of airway trees using multiple hypothesis tracking and template matching},
  author={Selvan, Raghavendra and Petersen, Jens and Pedersen, Jesper H and de Bruijne, Marleen},
  journal={arXiv preprint arXiv:1611.08131},
  year={2016}
}

@article{dorkenwald2023cave,
  title={CAVE: Connectome annotation versioning engine},
  author={Dorkenwald, Sven and Schneider-Mizell, Casey M and Brittain, Derrick and Halageri, Akhilesh and Jordan, Chris and Kemnitz, Nico and Castro, Manual A and Silversmith, William and Maitin-Shephard, Jeremy and Troidl, Jakob and others},
  journal={bioRxiv},
  year={2023},
  publisher={Cold Spring Harbor Laboratory Preprints}
}

@article{dorkenwald2024neuronal,
  title={Neuronal wiring diagram of an adult brain},
  author={Dorkenwald, Sven and Matsliah, Arie and Sterling, Amy R and Schlegel, Philipp and Yu, Szi-Chieh and McKellar, Claire E and Lin, Albert and Costa, Marta and Eichler, Katharina and Yin, Yijie and others},
  journal={Nature},
  volume={634},
  number={8032},
  pages={124--138},
  year={2024},
  publisher={Nature Publishing Group UK London}
}

@article{antiga2008image,
  title={An image-based modeling framework for patient-specific computational hemodynamics},
  author={Antiga, Luca and Piccinelli, Marina and Botti, Lorenzo and Ene-Iordache, Bogdan and Remuzzi, Andrea and Steinman, David A},
  journal={Medical \& biological engineering \& computing},
  volume={46},
  pages={1097--1112},
  year={2008},
  publisher={Springer}
}

@article{olufsen2000numerical,
  title={Numerical simulation and experimental validation of blood flow in arteries with structured-tree outflow conditions},
  author={Olufsen, Mette S and Peskin, Charles S and Kim, Won Yong and Pedersen, Erik M and Nadim, Ali and Larsen, Jesper},
  journal={Annals of biomedical engineering},
  volume={28},
  pages={1281--1299},
  year={2000},
  publisher={Springer}
}

@article{bullitt2003measuring,
  title={Measuring tortuosity of the intracerebral vasculature from MRA images},
  author={Bullitt, Elizabeth and Gerig, Guido and Pizer, Stephen M and Lin, Weili and Aylward, Stephen R},
  journal={IEEE transactions on medical imaging},
  volume={22},
  number={9},
  pages={1163--1171},
  year={2003},
  publisher={IEEE}
}

@article{krissian2000model,
  title={Model-based detection of tubular structures in 3D images},
  author={Krissian, Karl and Malandain, Gr{\'e}goire and Ayache, Nicholas and Vaillant, R{\'e}gis and Trousset, Yves},
  journal={Computer vision and image understanding},
  volume={80},
  number={2},
  pages={130--171},
  year={2000},
  publisher={Elsevier}
}

@article{Lee1994,
  author    = {Lee, T.C. and Kashyap, R.L. and Chu, C.N.},
  title     = {Building Skeleton Models via 3-D Medial Surface Axis Thinning Algorithms},
  journal   = {CVGIP: Graphical Models and Image Processing},
  volume    = {56},
  number    = {6},
  pages     = {462--478},
  year      = {1994},
  doi       = {10.1006/cgip.1994.1042}
}

@inproceedings{Amenta2001,
  author    = {Amenta, Nina and Choi, Sunghee and Kolluri, Ravi Krishna},
  title     = {The Power Crust},
  booktitle = {Proceedings of the sixth ACM symposium on Solid modeling and applications},
  pages     = {249--266},
  year      = {2001},
  doi       = {10.1145/376957.376986}
}

@article{Tagliasacchi2012,
  author    = {Tagliasacchi, Andrea and Alhashim, Ibraheem and Olson, Matt and Zhang, Hao},
  title     = {Mean Curvature Skeletons},
  journal   = {Computer Graphics Forum},
  volume    = {31},
  number    = {5},
  pages     = {1735--1744},
  year      = {2012},
  doi       = {10.1111/j.1467-8659.2012.03178.x}
}

@article{Abeysinghe2009Interactive,
  author = {Abeysinghe, Sasakthi S. and Ju, Tao},
  title = {Interactive skeletonization of intensity volumes},
  journal = {Vis Comput},
  volume = {25},
  pages = {627--635},
  year = {2009},
  publisher = {Springer-Verlag},
  doi = {10.1007/s00371-009-0325-5}
}

@article{BARBIERI201623,
 title = "An interactive editor for curve-skeletons: SkeletonLab",
 journal = "Computers \& Graphics",
 volume = "60",
 pages = "23 - 33",
 year = "2016",
 issn = "0097-8493",
 doi = "https://doi.org/10.1016/j.cag.2016.08.002",
 url = "http://www.sciencedirect.com/science/article/pii/S0097849316300905",
 author = "Simone Barbieri and Pietro Meloni and Francesco Usai and L. Davide Spano and Riccardo Scateni"
}

@article{Kruskal1956,
  author = {Kruskal, J. B.},
  title = {On the shortest spanning subtree of a graph and the traveling salesman problem},
  journal = {Proceedings of the American Mathematical Society},
  volume = {7},
  number = {1},
  pages = {48--50},
  year = {1956},
  doi = {10.1090/S0002-9939-1956-0078686-7}
}

@article{Prim1957,
  author = {Prim, R. C.},
  title = {Shortest connection networks And some generalizations},
  journal = {Bell System Technical Journal},
  volume = {36},
  number = {6},
  pages = {1389--1401},
  year = {1957},
  doi = {10.1002/j.1538-7305.1957.tb01515.x}
}

@article{Delaunay1934,
  author = {Delaunay, B.},
  title = {Sur la sphère vide},
  journal = {Izvestia Akademii Nauk SSSR, Otdelenie Matematicheskikh i Estestvennykh Nauk},
  number = {6},
  pages = {793--800},
  year = {1934}
}

@article{Dijkstra1959,
  author = {Dijkstra, E. W.},
  title = {A note on two problems in connexion with graphs},
  journal = {Numerische Mathematik},
  volume = {1},
  number = {1},
  pages = {269--271},
  year = {1959},
  doi = {10.1007/BF01386390}
}

@article{Hart1968,
  author = {Hart, P. E. and Nilsson, N. J. and Raphael, B.},
  title = {A Formal Basis for the Heuristic Determination of Minimum Cost Paths},
  journal = {IEEE Transactions on Systems Science and Cybernetics},
  volume = {SSC-4},
  number = {2},
  pages = {100--107},
  year = {1968},
  doi = {10.1109/TSSC.1968.300136}
}

@inproceedings{Ogniewicz1992Voronoi,
  author = {Ogniewicz, R. and Ilg, M.},
  title = {Voronoi skeletons: theory and applications},
  booktitle = {Proceedings of the IEEE Computer Society Conference on Computer Vision and Pattern Recognition (CVPR)},
  pages = {63--69},
  year = {1992},
  doi = {10.1109/CVPR.1992.223226}
}

@article{Dobbs2024Grapevine,
  title={Accurate 3D Grapevine Structure Extraction from High-Resolution Point Clouds},
  author={Dobbs, Harry and Peat, Casey and Batchelor, Oliver and Atlas, James and Green, Richard},
  journal={arXiv preprint arXiv:2502.20417},
  year={2024},
  eprint={2502.20417},
  archivePrefix={arXiv},
  primaryClass={cs.CV}
}

@article{zeng2021toporoot,
  title={TopoRoot: a method for computing hierarchy and fine-grained traits of maize roots from 3D imaging},
  author={Zeng, Dan and Li, Mao and Jiang, Ni and Ju, Yiwen and Schreiber, Hannah and Chambers, Erin and Letscher, David and Ju, Tao and Topp, Christopher N},
  journal={Plant methods},
  volume={17},
  pages={1--17},
  year={2021},
  publisher={Springer}
}

@article{jiang2019three,
  title={Three-dimensional time-lapse analysis reveals multiscale relationships in maize root systems with contrasting architectures},
  author={Jiang, Ni and Floro, Eric and Bray, Adam L and Laws, Benjamin and Duncan, Keith E and Topp, Christopher N},
  journal={The Plant Cell},
  volume={31},
  number={8},
  pages={1708--1722},
  year={2019},
  publisher={American Society of Plant Biologists}
}

@article{raumonen2013fast,
  title={Fast automatic precision tree models from terrestrial laser scanner data},
  author={Raumonen, Pasi and Kaasalainen, Mikko and {\AA}kerblom, Markku and Kaasalainen, Sanna and Kaartinen, Harri and Vastaranta, Mikko and Holopainen, Markus and Disney, Mathias and Lewis, Philip},
  journal={Remote Sensing},
  volume={5},
  number={2},
  pages={491--520},
  year={2013},
  publisher={Molecular Diversity Preservation International (MDPI)}
}

@article{siddiqui2021genetics,
  title={Genetics and genomics of root system variation in adaptation to drought stress in cereal crops},
  author={Siddiqui, Md Nurealam and L{\'e}on, Jens and Naz, Ali A and Ballvora, Agim},
  journal={Journal of Experimental Botany},
  volume={72},
  number={4},
  pages={1007--1019},
  year={2021},
  publisher={Oxford University Press UK}
}

@article{bucksch2014image,
  title={Image-based high-throughput field phenotyping of crop roots},
  author={Bucksch, Alexander and Burridge, James and York, Larry M and Das, Abhiram and Nord, Eric and Weitz, Joshua S and Lynch, Jonathan P},
  journal={Plant physiology},
  volume={166},
  number={2},
  pages={470--486},
  year={2014},
  publisher={American Society of Plant Biologists}
}

@inproceedings{ghaffari2015automatic,
title = {Automatic and patient-specific reconstruction of the cerebral vasculature, CSF spaces and parenchyma for hemodynamic assessment of vascular pathologies},
author = {Ghaffari, Mahsa and Hsu, Chih-Yang and Schneller, Ben and Zhou, Joe and Alaraj, Ali and Linninger, Andreas},
year = {2015},
month = {05},
pages = {},
booktitle= {CEREBROVASCULAR DISEASES},
doi = {10.13140/RG.2.1.3400.3043}
}

@article{gemci2008computational,
title = {Computational model of airflow in upper 17 generations of human respiratory tract},
author = {Gemci, Tevfik and Ponyavin, Valery and Chen, Yitung and Chen, H and Collins, Richard},
year = {2008},
month = {02},
pages = {2047-54},
volume = {41},
journal = {Journal of biomechanics},
doi = {10.1016/j.jbiomech.2007.12.019}
}

@article{Zhang2024NatMethods_CAR,
  author    = {Zhang, L. and Huang, L. and Yuan, Z. and Li, A. and Wang, Y. and Liu, S. and Yang, Q. and Gong, H. and Peng, H.},
  title     = {Collaborative augmented reconstruction of {3D} neuron morphology in mouse and human brains},
  journal   = {Nature Methods},
  year      = {2024},
  volume    = {21},
  number    = {10},
  pages     = {1936--1946},
  month     = {oct},
  doi       = {10.1038/s41592-024-02401-8},
  publisher = {{Nature Publishing Group US}}
}

\end{document}